# Extra Loss-free Non-Hermitian Engineered Single Mode Laser Systems

Mohammad H. Teimourpour[1], Hamed Dalir[2], Elham Heidari[3], Mario Miscuglio[1], Ray T. Chen[3], Demetrios N. Christodoulides[4], Volker J. Sorger[1,*]

1. Department of Electrical and Computer Engineering, George Washington University, 800 22nd Street, Science and Engineering Hall, Washington, DC 20052, USA

2. Omega Optics, Inc., 8500 Shoal Creek Blvd., Building 4, Suite 200, Austin, TX 78757, USA

3. Department of Electrical and Computer Engineering, The University of Texas at Austin, 10100 Burnet Rd., MER 160, Austin, TX 78758, USA

4. College of Optics & Photonics – CREOL, University of Central Florida, Orlando, Florida 32816, USA

*Email: sorger@gwu.edu

**Abstract:** In a laser system non-Hermitian methods such as Parity-Time (PT) Symmetry and Supersymmetry (SUSY) have shown and demonstrated the ability to suppress unwanted lasing modes and, thus, achieved single mode lasing operation through the addition of lossy passive elements. While these approaches enable laser engineering versatility, they rely on the drawback of adding optical losses to a system tasked to produce single mode gain. Unlike PT and SUSY lasers, here we show an extra loss-free non-Hermitian laser engineering approach to realize single mode lasing operation for the first time. By selectively enhancing the fundamental mode's quality factor, we obtain single mode operation with higher output power per cavity since all cavities in this system contribute to the laser output, in contrast to other non-Hermitian approaches. Furthermore, we show that this approach interestingly allows reducing the number of to-be-designed cavities in super-partner array as compared with, for example, the SUSY approach, thus leading to reduced design complexity upon coupled cavity scale up of laser arrays. In summary, the ability to engineer coupled laser systems where each laser cavity contributes to coherent light amplification opens up a new degree of laser-design freedom leading to increased device performance and simultaneous reduced design and fabrication complexity.

**Introduction:**

Non-Hermitian engineering of photonic systems for achieving certain functionalities has been a subject of passionate investigations in the past decade [1-12]. This interest was sparked by many applications such as single mode lasing [13-20], lasing dynamics control [21-23] and saturable absorber [24] to just mention a few [25].

Since the early days of the invention of the laser, a considerable number of approaches has been proposed to provide high-power laser systems for a variety of applications such as optical space communication, Light Detection and Ranging (LIDAR), spectroscopy and other purposes [26-40]. Among these methods, single mode lasing based on non-Hermitian engineering of the laser systems attracted considerable attention and different scenarios based on Parity Time (PT) symmetry [13-16] and Supersymmetry (SUSY) has been proposed so far [17-20]. In all of these approaches unwanted lasing modes has been suppressed via adding extra lossy elements in system in such a way that out-of-phase modes experience greater attenuation and higher lasing threshold compared to in-phase mode. These approaches have been studied both theoretically and experimentally [13-20], however, one of the main disadvantages of these methods is their efficiency. That is, in non-Hermitian laser engineering, the laser system consists of a number of distributed laser cavities, some of which contribute in laser emission power where others are passive lossy elements without any involvement in the output power. Thus, a large number of these laser cavities do actually not contribute to the laser's output power. It is therefore, that the laser output of such 'classical' non-Hermitian system scales sub-linear with the number of laser cavities. These inefficiency impacts all aspects of laser engineering to include real-estate footprint and slope efficiency to name a few [26-29].

To tackle this issue, here we show and discuss a different approach, in which, instead of suppressing the Quality factors ($Q$) of many out-of-phase modes, we judiciously design the system to enhance the $Q$ of the fundamental mode. The advantages of this method can be summarized as follows (1) single mode operation for the system with all cavity (lasing elements) contributing to output emission, (2) higher output power per number of distributed laser cavities, since the introduction of loss in the system is not required, (3) the laser design allows for a wider design geometry (i.e. reduced design complexity) corresponding to a more straight-forward laser design and fabrication process with a possible higher yield.

In this work we explore the fundamental properties of this novel non-Hermitian laser design by investigating (1) through (3). In this manuscript we start by introducing the $Q$-enhancing approach via Hamiltonian description for the simplest system of a photonic molecule (PM) [41] laser coupled to a partner cavity. In the second section, we have studied the laser dynamics and show the single mode in-phase lasing operation. Finally, we have studied some possible single mode laser arrays and compared their output power per cavity with their SUSY counterparts.

**Theoretical model:**

In the semiconductor lasers and in general all laser types, one cannot increase the pumping power to any given laser resonator to desired high value to achieve higher laser output power and expect the system remains stable [26-29]. There are several effects such as waste heat, gain saturation, and filamentation that may cause laser instability at very high pumping regime [26-29]. One approach to overcome this issue is to use laser arrays instead of a single laser element. Laser arrays are composed of several mutually coupled laser cavities similar to the same idea of coupled antenna arrays in microwave engineering. The main challenge however, for a laser array is to achieve in-phase lasing operation where each supermode of the array compete in reaching the lasing threshold. To overcome this multimode oscillation several methods have been proposed including non-Hermitian SUSY [17-20]. Here, we propose a new approach based on non-Hermitian Q-enhancing method.

To understand the *Q*-enhancing approach we start by the simplest nonlimiting system consisting of two identical coupled laser cavities at resonance frequency $\omega_0$, i.e. a PM, and this system is weakly coupled to a partner cavity supporting resonance at $\Omega = \omega_0 + \kappa$. Here $\kappa$ represents the coupling coefficient between identical laser cavities and $J$ is the weak coupling coefficient between PM and the partner cavity, as illustrated schematically in Fig. 1(A). Note, that for simplicity we assume that all cavities are single mode. In general, a laser cavity has several modes, however, it is possible to engineer a resonator's free spectral range (FSR) to achieve a large separation between modes of the cavity. For instance, in the case of a micro-ring/disk resonator, FSR can be easily increase by reducing the diameter of the cavity. In return, this implies larger mismatch between optical modes of two identical optical cavity. In other words, one can ignore the coupling between these modes and the cavity can be considered single mode in desired frequency window.

The system's behavior, within the framework of coupled mode theory before reaching to the lasing threshold where the system is linear, can be described by the following Hamiltonian $[16 - 18]$:

$$\widehat{H} = \begin{pmatrix} \omega_0 + i(g - \gamma) & \kappa & J \\ \kappa & \omega_0 + i(g - \gamma) & J \\ J & J & \Omega + i(g - \gamma) \end{pmatrix} \quad (1)$$

here, $g$ and $\gamma$ correspond to the gain and loss of cavities. In this system we assume that the coupling between partner cavity and the PM is weak enough in such a way that super modes of the PM remain intact, this can be realized under the assumption *of* $J \ll \kappa$:

$$\widehat{H} \approx \begin{pmatrix} \omega_0 + \kappa + i(g - \gamma) & 0 & \sqrt{2}J \\ 0 & \omega_0 - \kappa + i(g - \gamma) & \sqrt{2}J \\ \sqrt{2}J & \sqrt{2}J & \Omega + i(g - \gamma) \end{pmatrix} \quad (2)$$

The eigenvalues of this system are $\Omega_\pm = \omega_0 + \kappa \pm \sqrt{2}J + i(g - \gamma)$ and $\Omega_0 = \omega_0 - \kappa + i(g - \gamma)$. Here, mode $\Omega_0$ corresponds to the out-of-phase lasing mode which is not desired. The main idea here is to enhance the $Q$ factor of in-phase mode which implies a lower lasing threshold as compared to the out-of-phase modes. To do this, we introduce an extra gain $\Delta g$ at the partner cavity of the system and the effective Hamiltonian $\widehat{H}_{eff}$ of the system then reads:

$$\widehat{H}_{eff} \approx \begin{pmatrix} \omega_0 + \kappa + i(g - \gamma) & \sqrt{2}J \\ \sqrt{2}J & \Omega + i(g + \Delta g - \gamma) \end{pmatrix} \quad (3)$$

where the interaction between mismatched modes $\Omega_0$ and $\Omega$ is ignored, see Fig. 1(B).

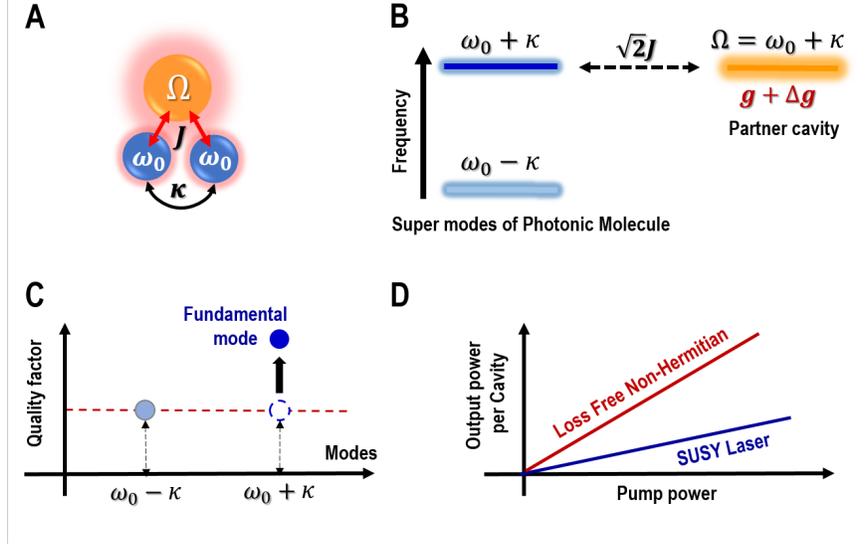

**Fig. 1:** (A) Schematic representation of two identical coupled laser cavities, PM, weakly coupled to another partner cavity at resonance $\Omega$. (B) As a result of mutual and identical coupling between partner cavity and PM, partners mode $\Omega = \omega_0 + \kappa$ is coupled to the in-phase mode ($\omega_0 + \kappa$), however, its coupling to the out-of-phase mode $\omega_0 - \kappa$ can be ignored since they are mismatch. (C) $Q$-factor of the in-phase mode (fundamental mode) increases as a result of extra gain ($\Delta g$) added to the partner cavity. One of the main advantages of extra-loss free non-Hermitian engineering of the laser system compared to other approaches such as SUSY is a significant increase in the output power per cavity in the laser array, this is shown schematically in (D), and quantitatively in Table 1.

The eigenvalues of this effective system read: $\omega_\pm = \omega_0 + \kappa + i\left(g + \frac{\Delta g}{2} - \gamma\right) \pm i\sqrt{\left(\frac{\Delta g}{2}\right)^2 - 2J^2}$. Clearly, for $\Delta g < 2\sqrt{2}J$ the real part of $\omega_\pm$ which corresponds to the lasing frequency varies as a function of input gain, and this regime is not desired. On the other hand, the system lives in the broken phase if $\Delta g > 2\sqrt{2}J$ which corresponds to the $Q$-enhanced in-phase lasing mode with lowest lasing threshold. This can be seen better for special case of $\Delta g \gg 2\sqrt{2}J$ in which $\omega_+ \approx \omega_0 + \kappa + i(g + \Delta g - \gamma)$ and $\omega_- \approx \omega_0 + \kappa + i(g - \gamma)$. The other two modes: $\omega_-$ and $\Omega_0$ have lower $Q$ compared to $\omega_+$ which corresponds to the in-phase mode, see Fig. 1(C). Noting that for the case of $J \sim \kappa$ in which partner cavity is not weakly coupled to the PM, in-phase mode of the PM alters dramatically which corresponds to mismatching and weak interaction with partner cavity which implies weaker Q-enhancement compared to the case of perturbative coupling regime i.e. $J \ll \kappa$.

In the next section we study the laser dynamics of this system. We find that output power per cavity for this system is higher than that of its SUSY counterpart, this is schematically illustrated in Fig. 1(D) and quantitatively validated in Table 1.

To illustrate $Q$-enhancing, we studied the imaginary parts of eigen-frequencies of the system for different gain $g$ and $\Delta g$, see Fig. 2(A). The horizontal red sheet illustrated in Fig. 2 (A) represents the lasing threshold at which $Im(\omega_\pm) = 0$. In Fig. 2(B), two distinguished domains are illustrated: regime $g < \gamma$ in which for $\Delta g > \Delta g_{th} = 2(g - \gamma)$ in-phase mode is the only lasing mode and out-of-phase mode remains below the lasing threshold. This regime where the fundamental mode lases but high-order modes not yet would be an ideal operating range of any laser system. However, for $g > \gamma$ system always lives in the out of phase operation regime.

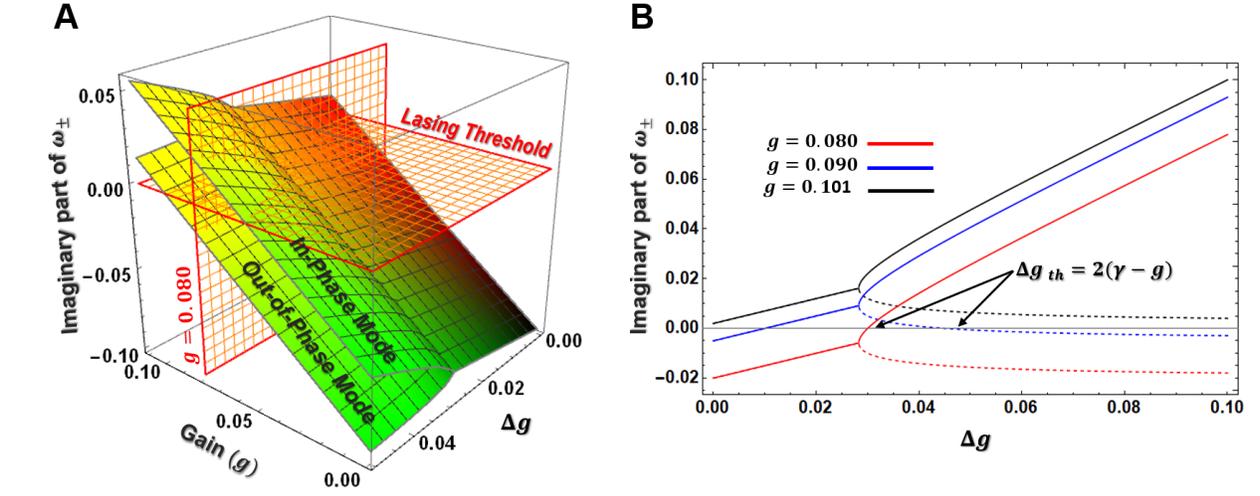

**Fig. 2:** Imaginary part of $\omega_\pm$ as a function of gain $g$ and its gradient $\Delta g$ for PM laser illustrated in Fig. 1: (A). horizontal (red mesh) sheet represents the lasing threshold. Clearly, the in-phase mode reaches the lasing threshold before out-of-phase mode. Here, for both cases loss is assumed to be $\gamma = 0.1$ and coupling coefficients are assumed to be as follows $\kappa = 1$ and $J = 0.01$. Nothing that $\kappa$ is normalized to unity in order to consider a general case and consequently all other parameters (including gain and loss) are normalized with respect to it. (B) For $g < \gamma$ in phase lasing threshold is at $\Delta g_{th} = 2(g - \gamma)$ in which out of phase is always below lasing threshold. On the other hand, for $g \geq \gamma$ system always lase in out of phase regime. Vertical plane in A at $g = 0.08$ corresponds to the black curve in B, the other two curves in B also correspond to two planes in A (not shown here) for given gain values $g$.

**Laser dynamics:**

In previous section we studied the static behavior of the system by describing the main concept of Q-enhancing. In what follows we investigate the dynamics of the system through solving the following system of coupled governing equations [16-18]:

$$\frac{\partial E_i}{\partial t} = \left(i\omega_i + \frac{g_i}{1+\alpha|E_i|^2} - \gamma_i\right) E_i + i\kappa_i(E_{i-1} + E_{i+1}) \quad (4)$$

here, $E_i$ represents the electric field in the $i^{th}$ laser cavity, $\omega_i$ is its lasing frequency and $\alpha = 1$ accounts for normalized gain saturation coefficient, $t$ is the normalized time and finally $g_i$ and $\gamma_i$ are gain and loss of the cavity, respectively. Noting that here, the normalized time $t = \kappa \cdot \tau$ is related to real time $\tau$ through the coupling coefficient $\kappa$.

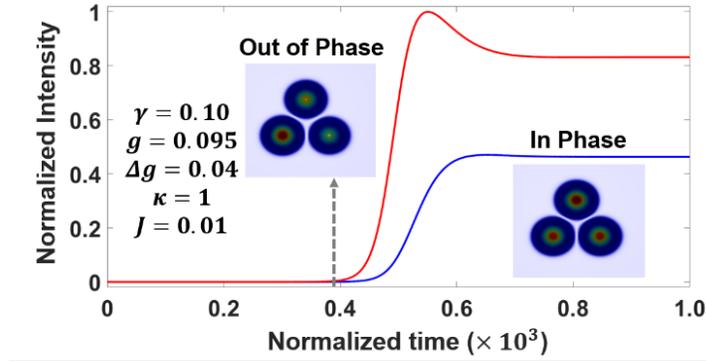

**Fig. 3:** Nonlinear dynamics of optical fields in laser molecule coupled to a partner cavity shown in Fig. 1. Here we also depict the temporal snapshots of the lasing pattern $E_i$ where steady-state single-mode emission can be observed after a transient period. System's parameters are described in the figure. Red and blue curves corresponds to the light intensities in the partner laser and PM cavities, respectively. Noting that the laser output powers are normalized to the maximum output intensity, in other words, peak intensity is considered to be equal to one. The same normalization performed in the rest of the paper.

In Fig. 3 we consider the system presented in Fig. 1 where in-phase single mode operation is found as illustrated by temporal snapshots of the electric field in cavities. This confirms that $Q$-enhancing approach can be used to achieve single mode operation in laser arrays. In the next section we study some of these laser arrays.

**Implementation to an array of coupled lasers:**

Next we utilize the above lasing cavity system which does not required an addition of extra lossy passive cavities to realize single mode laser action, but instead uses enhancement of the $Q$-factor of the fundamental in-phase laser mode to explore and study laser arrays. This is interesting because the newly gained degrees of design freedom enable different geometrical laser-design options. In fact, this approach bears significant universality, since such an approach can be applied to, essentially, any laser array system such as microring resonators [30], photonic crystals [31] and Vertical Cavity Surface Emitting Lasers (VCSELs) [36,40,].

While laser arrays bear a number of design and practical utilization advantages, there are fundamental challenges as well, one of which this newly introduced design methodology is able to address, namely the issue of multi-mode lasing leading up to lasing instability. Here we find and argue that rather than lowering the $Q$ of cavities of the system by adding loss, the same principle should work by reducing the loss or enhancing-$Q$. In addition to this, we are also interested in exploring arrays of lasers with different arrangements. To illustrate the array versatility and comparison to SUSY systems, in what follows we present examples of possible arrays to achieve in-phase single mode operation though presented loss free non-Hermitian method.

In general, there are many different possibilities to construct a laser array: one dimensional (1D) and two dimensional (2D) arrangements [26-29]. However, 2D arrays are especially appealing for vertical emitting laser platforms such as VCSELs and Photonic crystals laser arrays in which in contrary to conventional edge-emitting semiconductor lasers, their emission should be controlled in the perpendicular direction. This type of laser arrays has been studied in considerable literatures including and more specifically for VCSELs in [43] in which many different possibilities for 2D VCSEL arrays has been proposed. Here, we just choose circular array arrangements as an example to show how our method can be applied to them to exploit in-phase lasing operation. In-phase operation of these array can plays an important role in optical systems such as LIDARs and optical sensing applications.

A circular array in which each laser cavity occupies vertex of a polygon is shown in Fig. 4(A). For such a system the in-phase mode corresponds to eigen-frequency $\Omega = \omega_0 + 2\kappa$, which is interestingly independent of $N$ (the number of cavities in circular array) [43]. To achieve single mode operation now we need to find a partner cavity exactly in the center of the circular array at resonance frequency $\Omega$. Considering additional gain in this central cavity assure in-phase single mode operation of the system.

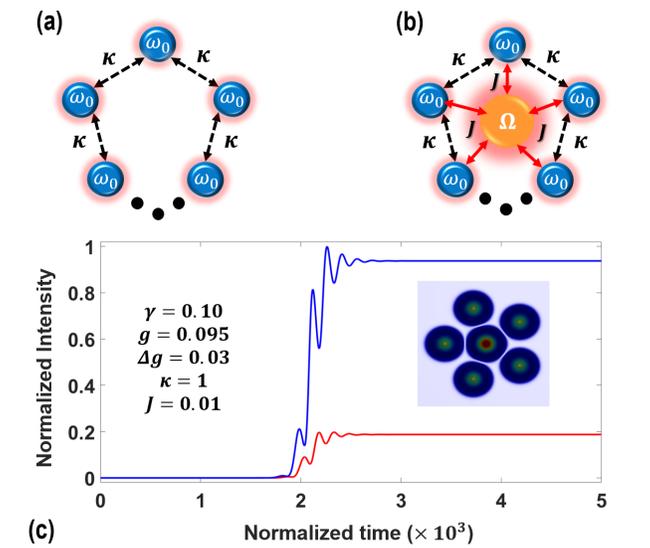

**Fig. 4:** (A) Circular array of N coupled cavities. (B) Centered polygonal array consists of a circular array coupled to a partner cavity at resonance $\Omega = \omega_0 + 2\kappa$ to achieve single mode operation. (C) Nonlinear dynamics of optical intensity $|E_i|^2$ in laser array for given parameters in inset. Blue and red curves corresponds to the light intensities in the partner laser and PM cavities, respectively.

**Comparison to the SUSY system:**

Laser design efficiency in terms of output power per cavity is an important aspect of any laser array system, here we compare our non-Hermitian engineering method with previously proposed SUSY lasers for different number of cavities. It is shown that for proposed extra loss-added free scenario, enjoying more gain than loss for every laser cavity plays a critical role in having higher efficiency (compared to SUSY) in converting the gain power into coherent radiation laser power.

We first start with simplest case which is $N = 3$ where $N$ is number of cavities in the system. SUSY laser arrays are based on suppression of unwanted modes of the system by introducing an evanescent coupling between a lossy super-partner and the main laser array [17,18]. In the case of $N = 3$ we have two active cavities, PM, which are coupled weakly to a passive super-partner at resonance $\Omega = \omega_0 - \kappa$, Fig. 5.

Total output laser power is obtained by solving the laser rate equations and then summing up all laser output intensities. In this section, to find radiated power per cavity we then divide the total output power by the number of active (only pumped) cavities in each system (Table 1).

For this illustration example, we also consider arrays with $N = 5$ and $N = 7$ and not even ones (e.g. $N = 4$ and $N = 6$) as they cannot be supported by the SUSY approach [17,18]. Refer to Figure 5 for details on the design. In [17,18] it is explained that for any given number of elements $N$ in an array there are infinite number of possibilities to generate the super-partner array as well as its coupling arrangement to the main array. Naturally, this sets up an optimization problem to be considered for engineering the laser. The SUSY super-partner for the illustrative cases in figure 5 are generated following the same method as described in [17,18]. The results indeed validate the anticipated higher output power per cavity for all cases engineered based on loss free non-Hermitian approach over their SUSY counterparts (Table 1). However, the order of magnitude of exceeding a 10-fold higher emission output is noteworthy, and somewhat non-expected, since the trivial ratio of more-cavities lasing is exceeded; for example for $N = 5$ the SUSY has 3 (blue) lasing cavities each producing $3 \times 0.03 \approx 0.1$ total power (in a.u. units) whilst this novel design introduced here produces $5 \times 0.42 = 2.1$ – a ratio of over 20 fold between these two approaches.

The reason for this improvement can be understood as a result of difference in systems architecture, since in the SUSY laser array the number of lasing elements is almost equal to the non-lasing passive elements which implies more power attenuation in the system. On the other hand, in the extra loss-free method of this work all cavities are lasing and none of them are passively lossy. It has to be also mentioned that in both of these scenarios there are infinite possibilities to engineer systems and optimization of these methods will be considered in future work. Interestingly, such cavity symmetry could be introduced into emerging nanostructures not just for light emitters but also for tailoring photo absorption with application in nanoscale solar-cells [44].

| Number of laser cavities ($N$) | Laser Output (a.u.) SUSY | Laser Output (a.u.) This Work | Improvement Factor |
|---|---|---|---|
| 3 | 0.03 | 0.10 | 3x |
| 5 | 0.03 | 0.42 | 14x |
| 7 | 0.03 | 0.26 | 9x |

**Table. 1:** Comparing the normalized output power per cavity for different number of cavities in laser arrays. One can see that, SUSY provides less power per cavity compared to presented approach in this work.

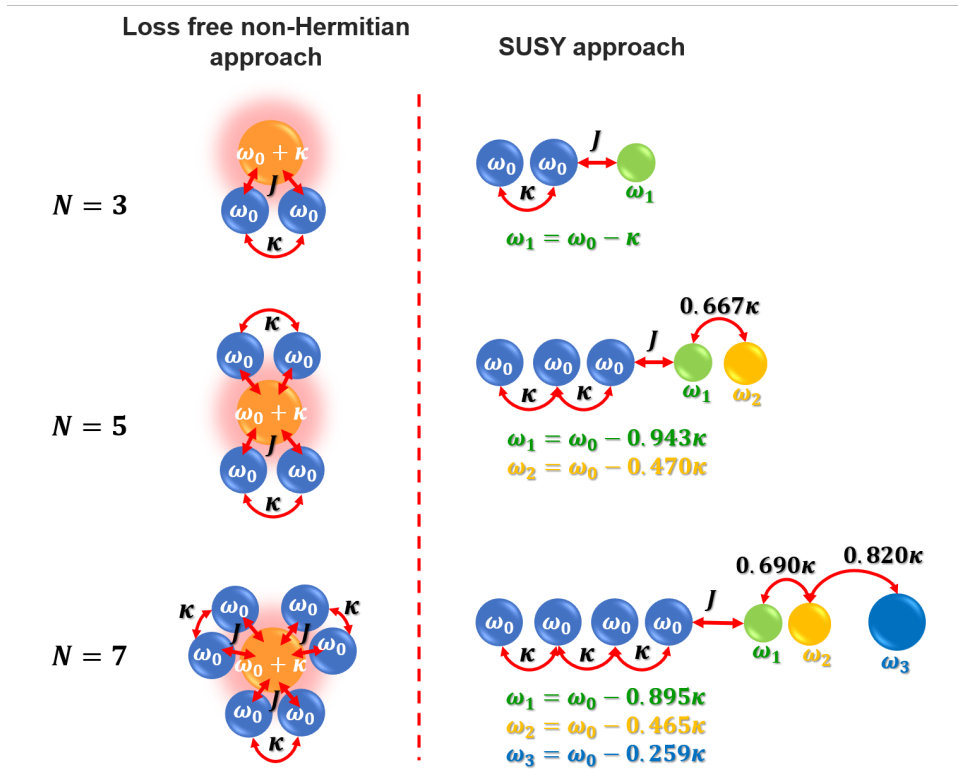

**Fig. 5:** Schematic of laser arrays consist of $N = 3$, 5 and 7 cavities: left column represents mutually coupled cavities to the central cavity based on our proposed approach in this work and their SUSY counterparts on the right panel. It must be mentioned that both above cases are just examples of all possible arrangements and one can consider many different architectures for these arrays. Here, all laser cavities are considered to have the same loss coefficient $\gamma = 0.1$ and $\kappa = 1$ and $J = 0.01$. Note, that here for $N = 3$ we consider the same total gain coefficient provided to all systems: for loss free engineered case it is $g_{total} = 0.30$ and for SUSY it is $g_{total} = 0.20$ which implies equal average power in every single cavity ($g_{avg} \approx 0.10$) for all cases. The same average power is considered for the other cases with $N = 5$ and 7.

## Conclusion remarks

Here we presented a novel method to achieve in-phase single mode operation in any laser array such as VCSELs and Photonic crystal lasers. This method is based on $Q$-enhancing of the fundamental supermode. Presented approach is then implemented in an array of coupled laser cavities. It is shown that the main advantages of extra loss free non-Hermitian approach over SUSY lasers are as follows: (1) In contrast to PT symmetric or SUSY single mode lasers where in which one or several cavities are passive without contributing in laser output power, here all cavities are lasing and none of them is passive, thus, higher output power per cavity are achieved. (2) Easier laser system engineering independent of complicated super partner arrays is a vital characteristic of our presented method which plays an important role in laser engineering.

**Methods**

Standard eigenvalue solver used for analyzing the linear eigenvalue problem of laser system at threshold. Runge Kutta 4 (RK4) algorithm used for solving systems of nonlinear coupled equations to investigate the laser dynamics.